\documentclass{amsart}

\usepackage{graphicx}
\usepackage{natbib}
\usepackage{amsmath,amssymb,amsthm}

\begin{document}

\title[A Bayesian Approach with Parametric Likelihood]{Using Data to Tune Nearshore Dynamics Models: 
A Bayesian Approach with Parametric Likelihood}

\author{Nusret Balci}
\address{Department of Mathematics, University of Arizona,
Tucson, AZ 85721, USA.}
\email{shankar@math.arizona.edu}

\author{Juan. M. Restrepo}
\address{Department of Mathematics, University of Arizona,
Tucson, AZ 85721, USA.}
\email{restrepo@math.arizona.edu}

\author{Shankar C. Venkataramani}
\address{Department of Mathematics, University of Arizona,
Tucson, AZ 85721, USA.}
\email{shankar@math.arizona.edu}

\begin{abstract}
We propose a modification of a  maximum likelihood procedure for tuning parameter values in models, based upon the comparison of their output to field data. Our methodology, which uses polynomial approximations of the sample space to increase the computational efficiency,  differs from similar Bayesian  estimation frameworks in the use of an alternative likelihood distribution, is shown to better address problems in which covariance information is lacking, than its more conventional counterpart.
  
 Lack of covariance information is a frequent challenge in large-scale geophysical estimation. 
 This is the case in the geophysical problem considered here. We use a nearshore model for long shore currents  and observational data of the same to show the contrast between both  maximum likelihood
methodologies. 

Beyond a methodological comparison, this study gives estimates of  parameter values for the bottom drag and surface forcing that make the particular model most consistent with data; furthermore, we also derive sensitivity estimates that provide useful insights regarding the estimation procedure as well as of the model itself.


\end{abstract}

\maketitle

\noindent{\em Submitted to Ocean Modeling}


\section{Introduction}

We have two principal goals in this work -- (1)  Introduce an alternative formulation to a maximum likelihood procedure for parameter estimation, which greatly improves estimates when covariance 
information of model and data is lacking; (2) Produce estimates of two important parameters critical to making model outcomes of longshore currents compatible with existing data. We will demonstrate the practicality and usefulness of our procedure by applying it to tune parameters in a model for nearshore dynamics using field data.

We are interested in parameter estimation for geophysical models which involve potentially complex models and a large number of observations.  A Bayesian framework  is natural for such geophysical problems: it is unrealistic to expect to completely determine ocean states via data alone.  Commonly used  {\it data assimilation} methods for dynamic geophysical models are based upon formulating a variance minimizer of the posterior distribution of model state, given observations (see \cite{wunschbook} for a review). The parameters are then inferred from the posterior distribution of model states. 
  
 For linear/Gaussian methods based upon least squares, it is prudent that we not declare the parameters as  state estimation variables, since the ensuing state estimation problem is
 typically highly  nonlinear and non-Gaussian. There are assimilation methods capable of dealing with nonlinearities, but these tend to be dimensionally-challenged: only capable of handling problems with a small number of effective dynamic state variables ({\it c.f.}, \cite{drifter}, \cite{ERA-I},  \cite{voss}, to name a few). In the data assimilation methods mentioned above the  posterior is written in terms of a likelihood that is informed by observations, and a prior which is instead informed by model outcomes. Fundamentally this presumes that models and data are error-laden and most critically, that these errors are very well known or estimated.

The parameter estimation method we use in this work is the same as the methods discussed above {\it if} we presume that the model is free of any errors.  Alternatively, it is estimation based upon maximum likelihood (see \cite{Loglike}). In its most basic form, one writes down a likelihood based upon knowledge of the statistics of the errors between models and data. One then uses sampling methods to find the most likely parameters. In this work we are  only working with  2 parameters so it is possible to circumvent the use of Monte Carlo and instead generate a table of  the likelihood function (plots of which  will be 
shown in this paper) in sample space. Once the parameter space is mapped out, it is then straightforward  to pick approximate maximum likelihood parameter combinations that lead to the best compatibility possible between model outcomes and observations. 
Whether one uses Monte Carlo or not, the computation of sample space is exceedingly expensive when geophysical models based on partial differential equations and/or many equations are involved. 
With the aim of improving the efficiency of  producing large number of model outcomes 
 engineering researchers have recently proposed 
using random-coefficient polynomials expansions of sample model outcomes ({\it cf.}, \cite{basicpc}). We will  adopt such a strategy here.  The model proxy will be  based upon a Polynomial Chaos expansion of the sample space ({\it cf.}, \cite{pcwiener}). Demonstrations of the
use of this expansion for the purpose of improving the efficiency of a Monte Carlo maximum likelihood parameter estimation in geophysical models are found 
in the works of \cite{polychaos1} and \cite{polychaos2}, and references contained therein.

The maximum likelihood method is applied to the estimation of 2 parameters critical to  nearshore longshore currents. We use  a  {\it vortex force formulation} for  the evolution of waves and currents in the nearshore (see \cite{mrl04},
 and \cite{lrm06}) which can capture these currents. The model 
was used in \cite{WULRM} to describe the evolution of rip currents.  Using this model
\cite{UMR} found that  longshore current outcomes  were most sensitively  dependent on the bottom drag force (and its parametrization), 
and the amplitude of the incoming waves (which are boundary conditions in the model). Since the drag force and the incoming wave forcing are such a critical part of a nearshore calculation using this wave/current model or some other model, 
it is essential to develop strategies to tune the parameter appropriately, particularly if the model is being used
to explore phenomena that are less familiar than the rip currents or longshore currents.

\cite{UMR}  also found that while different   bottom drag parametrizations
resulted in different longshore outcomes, it was often the case that
one could replicate qualitative and quantitative characteristics of the longshore currents with different models if the coefficients in the parametrization were chosen appropriately.
This suggests, in the setting considered, the type of parameterization is perhaps less important than precise tuning of the parameters for the model to accurately reproduce the physical outcome. We are thus motivated to test the capabilities of the simplest of drag parametrizations, the linear drag force model, but with an accurate tuning of the parameters through comparison with field data.

The data we use was collected in the field campaigns conducted in
Duck, North Carolina by Herbers, Elgar, and Guza in 1994 (see \cite{elgar1994}). The data
sheets and detailed information is readily available on the web\\
 {\tt frf.usace.army.mil/pub/Experiments/DUCK94/SPUV}.
We use the data collected in the bar region of
the sea bed topography, since this is the region where we observe
the  wave-induced strong longshore current which the vortex force model aims to  capture.

The paper is organized as follows. A summary of the model appears in Section \ref{model}. Our parameter estimation method using an alternative maximum likelihood formulation is described in Section~\ref{method}. We derive a sensitivity analysis estimate that is useful in the interpretation of model and observations. This analysis is presented in Section \ref{analysis}.
The outcomes of the test and physical interpretation of the results appear in Section \ref{outcomes}.

\section{The Wave/Current Interaction Model}
\label{model}
\begin{figure}
\centering
 \includegraphics[scale=0.35]{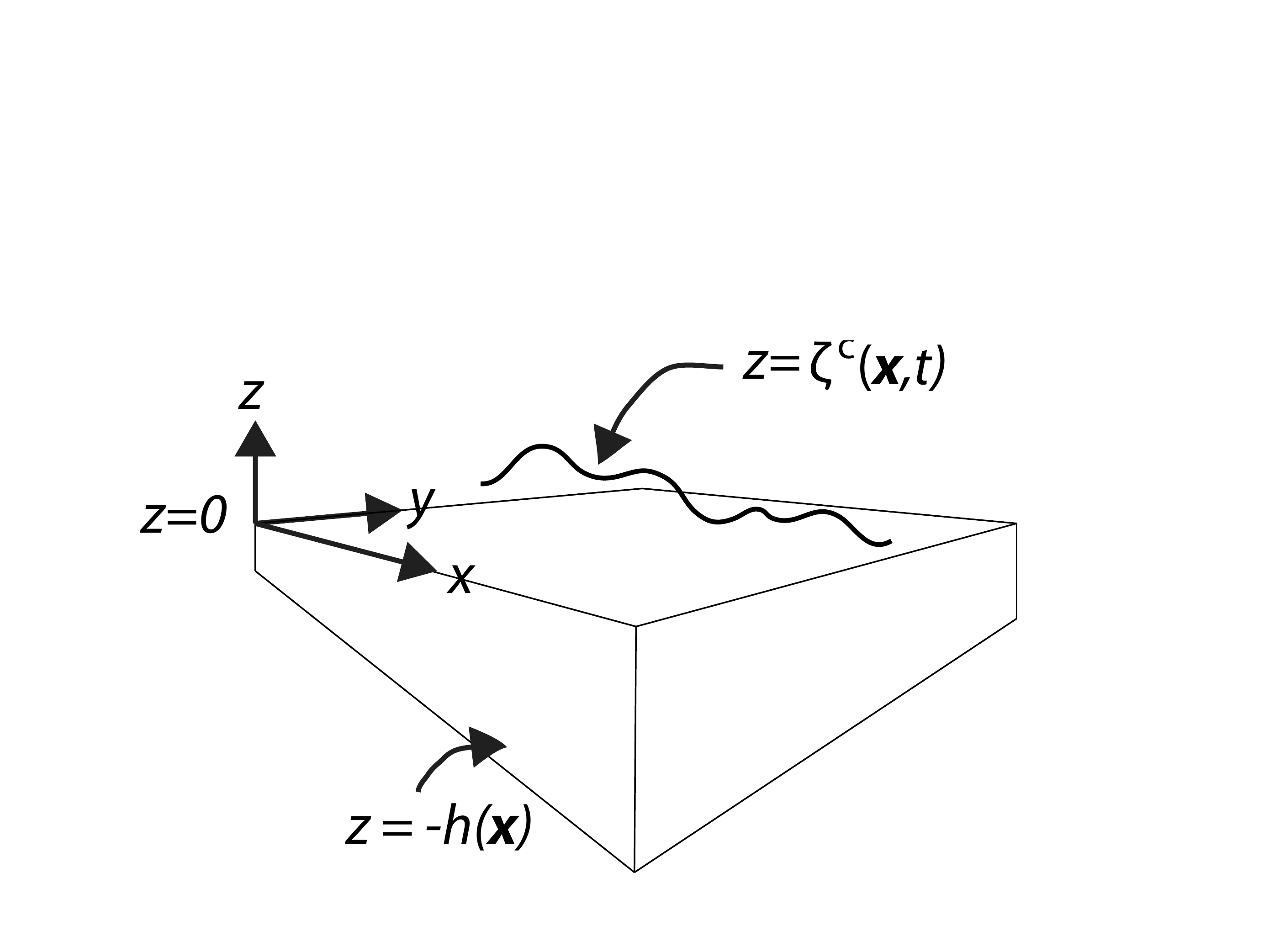}
\caption{Schematic of the nearshore environment.}
\label{fig:domain}
\end{figure}

The depth-averaged wave-current interaction model in \cite{mrl04} is specialized to the nearshore 
environment. See Figure \ref{fig:domain}.
The transverse coordinates of the domain will be denoted by ${\bf x}:=(x,y)$. The cross-shore coordinate is $x$ and increases away from the
beach. Time is denoted by $t \ge 0$. Differential operators depend only on ${\bf x}$ and $t$. The total water column depth is given by $H=h({\bf x}) + \zeta^c({\bf x},t)$,
where $h$ is the bottom topography and
$\zeta^c= \hat \zeta + \zeta$, is the composite sea elevation;  $\hat \zeta$ is the quasi-steady sea elevation adjustment,
$\hat{\zeta}=-A^{2}k/(2\sinh(2kH))$,
where $A$ is the wave amplitude and $k$ is the magnitude of the peak wavenumber. 
In the absence of wind forcing and for spatio-temporal scales much larger than those typical of the waves, the momentum equation reads
\begin{align}
\frac{\partial\mathbf{u}}{\partial t}+(\mathbf{u}\cdot\nabla)\mathbf{u}+g\nabla\zeta= & \mathbf{J}+\mathbf{B}-\mathbf{D}
+ {\bf S} + {\bf N},
\label{eq:momentum}
\end{align}
where $\mathbf{u}=\mathbf{u}(\mathbf{x},t)$ is the depth-averaged
transverse velocity, $g$ is the gravitational acceleration, 
 $\mathbf{J}$ is the vortex force due to waves
(see \cite{MR99}), $\mathbf{B}$ is the wave breaking acceleration,
and $\mathbf{D}$ is the momentum loss due to the bottom drag. 
${\bf S}$ is the wind stress, which will be taken to be zero in what follows, as
is ${\bf N}$, which represents   important dissipative effects from the subscale turbulent flow.

The vortex force term is defined as
\[
\mathbf{J}=-\mathbf{z}\times\mathbf{u}^{\mathrm{st}}\chi,
\]
where $\chi$ is the vorticity. 
We use a linear bottom drag formulation:
\[
{\mathbf \tau}=d \mathbf{u}.
\]
Then, the term on the right hand side of (\ref{eq:momentum}) is 
\[
\mathbf{D}=\frac{{\mathbf \tau}}{\rho H},
\]
where $\rho$ is the fluid density. 

The contribution to the flow due to the breaking waves has the form
\[
\mathbf{B}=\frac{\epsilon\mathbf{k}}{\rho H\sigma}.
\]
The function $\epsilon$ is arrived at by hydraulic jump theory. Taken from \cite{tgbreaking}),
it reads
\[
\epsilon=24\sqrt{\pi}\rho g\frac{B_{r}^{3}}{\gamma^{4}H^{5}}\frac{\sigma}{2\pi}A^{7},
\]
with $B_{r}=0.8$, $\gamma=0.4$. The 
peak wavenumber of the gravity wave field ${\bf k}$ and  $\sigma$  the wave frequency
obey the dispersion relation
\[
\sigma^{2}=gk\tanh(kH).
\]

The evolution of the water column height is given by the continuity equation
\begin{align}
\frac{\partial H}{\partial t}+\nabla\cdot [ H(\mathbf{u}+\mathbf{u}^{\mathrm{st}})] & =0,\label{eq:continuity}
\end{align}
where
\begin{equation}
\mathbf{u}^{\mathrm{st}}:=(u^{\mathrm{st}},v^{\mathrm{st}})=\frac{1}{\rho H}W\mathbf{k}.\label{eq:stokesdrift}
\end{equation}
is the Stokes drift velocity. The ray  equation
for the wave action
\[
W:=\frac{1}{2\sigma}\rho gA^{2}.
\]
 is given by
\begin{align}
\frac{\partial W}{\partial t}+\nabla\cdot(W\mathbf{c}) & =-\frac{\epsilon}{\sigma},\label{eq:waveaction}
\end{align}
where $\mathbf{c}$ is the group velocity,
given by the formula
\[
\mathbf{c}=\mathbf{u}+\frac{\sigma}{2k^{2}}\left(1+\frac{2kH}{\sinh(2kH)}\right).
\]
The conservation law for the
wavenumber reads
\begin{align}
\frac{\partial\mathbf{k}}{\partial t}+\mathbf{c}\cdot\nabla\mathbf{k} 
& =-(\mathbf{k}\cdot\nabla)\mathbf{u}-\frac{k\sigma}{(\sinh(2kH))}\nabla h.\label{eq:wavenumbers}
\end{align}

\subsection{Model Computations}

In the computations to be discussed subsequently  we  assumed that all of the fields were periodic in $y$.
At the near-shore coordinate $x=x_0$, where $x_0 = 100$ in the particular configuration used in the numerical simulations,
we imposed the condition $u = -u^{St}$ on the cross-shore 
component of the current velocity. In numerical
computations, $\mathbf{u}$ was relaxed towards this boundary conditions
over a layer. This layer, hugging the near-shore, $x=x_0$ side,  was selected so as to not affect the 
statistical comparisons and results 
in the region of interest. We imposed the homogeneous Neumann
boundary condition on $\zeta$ at $x=x_0$, and we nudged $\zeta$ toward zero
on the offshore boundary, at $x=L$, where $L = 600$ in the numerical simulations. 
Both $W$ and $\mathbf{k}$ were prescribed 
at $x=L$:  $W$ is prescribed so as to satisfy the offshore
wave amplitude parameter value. The wave number was chosen to make an
angle in the range 
$170$ to $190$ degrees, with respect to the shore-normal vector, its magnitude was
set once the frequency was set. Near the shore,
we imposed perfectly matched layer boundary conditions on  $W$, whereas 
 homogeneous
Neumann boundary condition were imposed on $\mathbf{k}$.

We used  data  from  the experiments conducted by Herbers,
Guza, and Elgar,  conducted in 1994, in Duck, North Carolina.
We will refer to the field data as the {\it Duck data.}
The Duck data also provides  information on the mean velocity, pressure, temperature, and depth. In addition, there is information on the peak frequency, and bottom topography.
We will be making use of sea elevation data as well as depth-averaged velocity data.

\begin{figure}
\centering
 \includegraphics[scale=1]{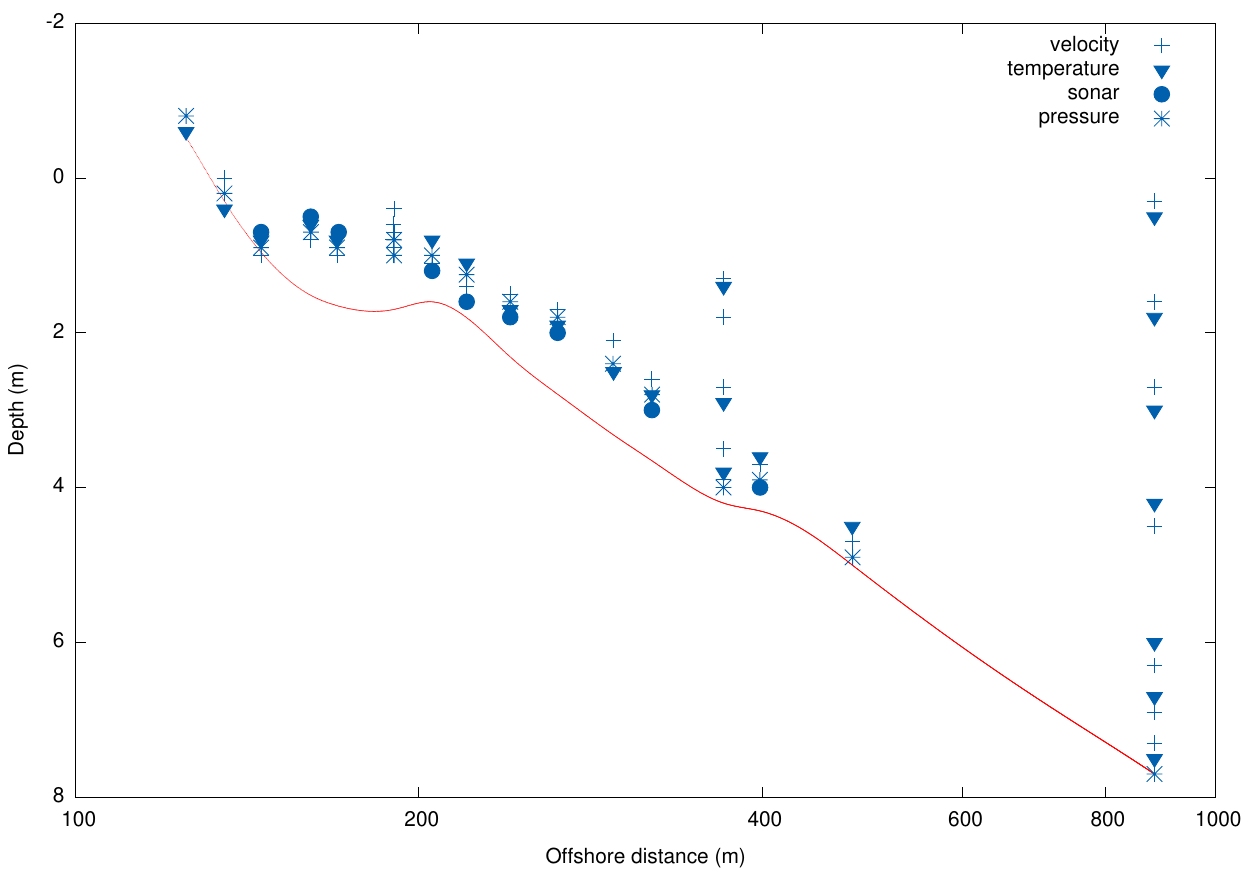}
\caption{The bottom topography, based upon the Duck data. Measurement locations
and the type of measurements are indicated.}
\label{fig:topography}
\end{figure}

The computational bottom topography, shown in Figure \ref{fig:topography},
was generated by joining their bathymetric information 
 using a cubic spline in the cross-shore direction. No 
 $y$ variation was assumed for the bottom topography.
The model was approximated using
second order finite-differences in space, and Heun quadrature in time.
The time steps were around $0.01$ s. 
The computational grid was uniform. The spatial grid
width in $y$-direction was about $4$ m ($61$ grid points are used),
 and in $x$-direction the grid width was about $1.95$ m in length and 257 grid points were used. 
The computational domain covered the cross-shore coordinates, $x$,
between $100$ m and $600$ m, and was $240$ m wide in the $y$-direction. 
The effective domain, which was largely
free of effects of the relaxation and the matching layer used in the
numerical computations, encompassed shore distances from $150$ m to $500$ m.

Among the things we know about the model, as applied to the longshore current problem
(see \cite{UMR}), 
is that  for high values of the drag parameter, the bottom drag force  is effectively in balance with  the breaking force, and the inertial effects are largely ignorable.  On the other hand, when the bottom drag parameter is small, the longshore current develops (non-stationary) instabilities.

\section{Bayesian Framework for Parameter Estimation}
\label{method}
The goal is to find estimates of the offshore wave boundary data  $a:=2A(L,t)$, where $A(x,y,t)$ is the wave amplitude
given by the relation $W=A\mathbf{k}$
 (we will refer to this
quantity as the {\it wave forcing}), and the bottom
{\it drag coefficient} $d$ that best agree with the data. In this study, $a$ is a boundary condition parameter and is time-independent. Specifically, we will create tables 
of the likelihood estimators for the wave forcing and the bottom drag. The parameter combinations that lead to the best agreement between model and data are the most likely from the table.

Based upon our understanding of the physics of the problem the 
parameter range for the wave forcing and  bottom drag were taken as
\begin{align}
a \in [a_0,a_1]=[0.4,1.2],  \quad d \in [d_0,d_1]=&[0.002,0.026],
 \label{eq.ranges} 
\end{align}
respectively.
 
 A fundamental assumption in the estimate (but not of the method itself), is that
 at the times $t_n$, $n=1,2,...,N$, when data is available,
 the statistical distribution of
 \[
 {\bf p}- G( {\bf q}),
 \]
 is normal. Here, ${\bf p} \in \mathbb{R}^m$ are field measurements,
 ${\bf q} \in \mathbb{R}^k$ is the  state vector. $G$ relates the measurements and the state vector.  A most 
common situation in  geophysical problems, such as the longshore problem,  is that
 $m \ll k$, and $T_N:=\min_{1 \le n \le N} (t_{n+1}-t_n)$ is small compared with
 $1/\sigma$. 
 An important assumption used here is that correlations in time are considerably shorter than $T_N$.
 
 In our specific nearshore example,   at time $t_n$, the dimension of ${\bf q}$ is $2$, times the number of space-grid locations (discounting
 for periodicity and boundary data). The measurement vector ${\bf p}$,  consists of 
  the time dependent sea elevation
 and depth-averaged velocities at measurement locations. $G$ is a projection matrix
 in this problem.  
 
 The argument
 of the likelihood shall measure the absolute weighted distance between data and model outcomes for  the sea elevation and velocities at measurement times $t_n$.  We will denote the model output at time $t_n$ as
   ${\mathcal M}_n$. Similarly, we will define the  data vector as ${\mathcal O}_n$. 
(The observation and the model vectors in the argument have the same dimension, thanks to the implied projection matrix).   
    Assuming Gaussianity in the likelihood,
 \begin{equation}
P(a,d\mid\mathcal{O}) \propto \prod_{n=1}^{N}\exp\left(-\frac12 [{\mathcal M}_{n}-\mathcal{O}_{n}]^\top R^{-1}_n [{\mathcal M}_{n}-\mathcal{O}_{n}] \right)
{\mathcal U} ([a_{0},a_{1}]){\mathcal U}([d_{0},d_{1}]),
\label{eq.pad}
\end{equation}
$R_n:=R(t_n)$ is a covariance matrix, and $\mathcal{U}[\alpha,\beta]$ denotes the uniform distribution over the range $[\alpha,\beta]$.
The first term in (\ref{eq.pad}) is the likelihood $P(\mathcal{O}|a,d)$. The likelihood is a product 
because we are assuming that the distributions of the measurement errors and the
model outcomes are independent in time.

Instead of inverting the data set to eke out the implicitly-defined model  parameters $a$ and $d$ is circumvented by making the following assumption: 
 the mode of the posterior probability density for $a$ and $d$ given observations is
 obtained when the argument of the   likelihood distribution is minimized.
Compared to a standard proposal for data assimilation, (\ref{eq.pad}) does not 
include a prior  informed by explicit model error independent of errors due to measurements. This is not to say that the model is error-free, but rather, that all
that can be discerned is discrepancies (differences) between model output and measurements.
In terms of parameter estimation methodology, the maximum likelihood approach 
described above does not differ in any significant way  from the one proposed by \cite{polychaos1}.   In the next section we will argue for  an alternative likelihood function, thus distinguishing our work from theirs.

\subsection{Alternative Bayesian Statement}
\label{abs}


The matrix $R_n$ provides a description of covariances among the different
components of the state vector, the degree of confidence in each of these, as well as
 scale/non-dimensionalization information for each of the state components. It is essential information that must be known in order to use (\ref{eq.pad}) for parameter estimation.

 If not supplied,  one has to turn to whatever model and observations are available in order to estimate the covariance matrix.
In fact, in geoscience applications it is often the case that the covariance is unknown  or very poorly constrained. This 
is the case in the longshore  problem.  Let
 ${\mathcal M}_n$ represent the state vector at time $t_n$, produced by the computer-generated model solution, for a given set of parameter values, $a,d$. The state vector
will consist of dynamic variables ({\it e.g.}, the transverse velocity and sea elevation) at spatial locations with offshore $x_1$, $x_2$, ..., $x_k$. The individual state vector component, at time $n$ and location $x_j$, $j=1,..., k$,  will be denoted by $M_n(x_j):=M^j_n$. 
We will omit the superscript $j$, if it is clearly implied by the context.  $M_n$ is $y$-averaged (alongshore), and time averaged over times $t_{n-1}$ and $t_n$.
Because of the averaging, we can assume that the model outputs give accurate estimates for the mean longshore velocity for the given parameters at the locations $x_1,\ldots x_k$. 
 The individual component of the observation vector, measured
at location $x_j$, $j=1,2,...,k$, will be denoted by $O^j_n$ (again, we will omit the superscript $j$ unless it is not implied by the context).
The observations  $O_n$  are {\em not averaged in $y$ or time} and hence have significant variability. 

\begin{figure}
\centering
 \includegraphics[scale=0.5]{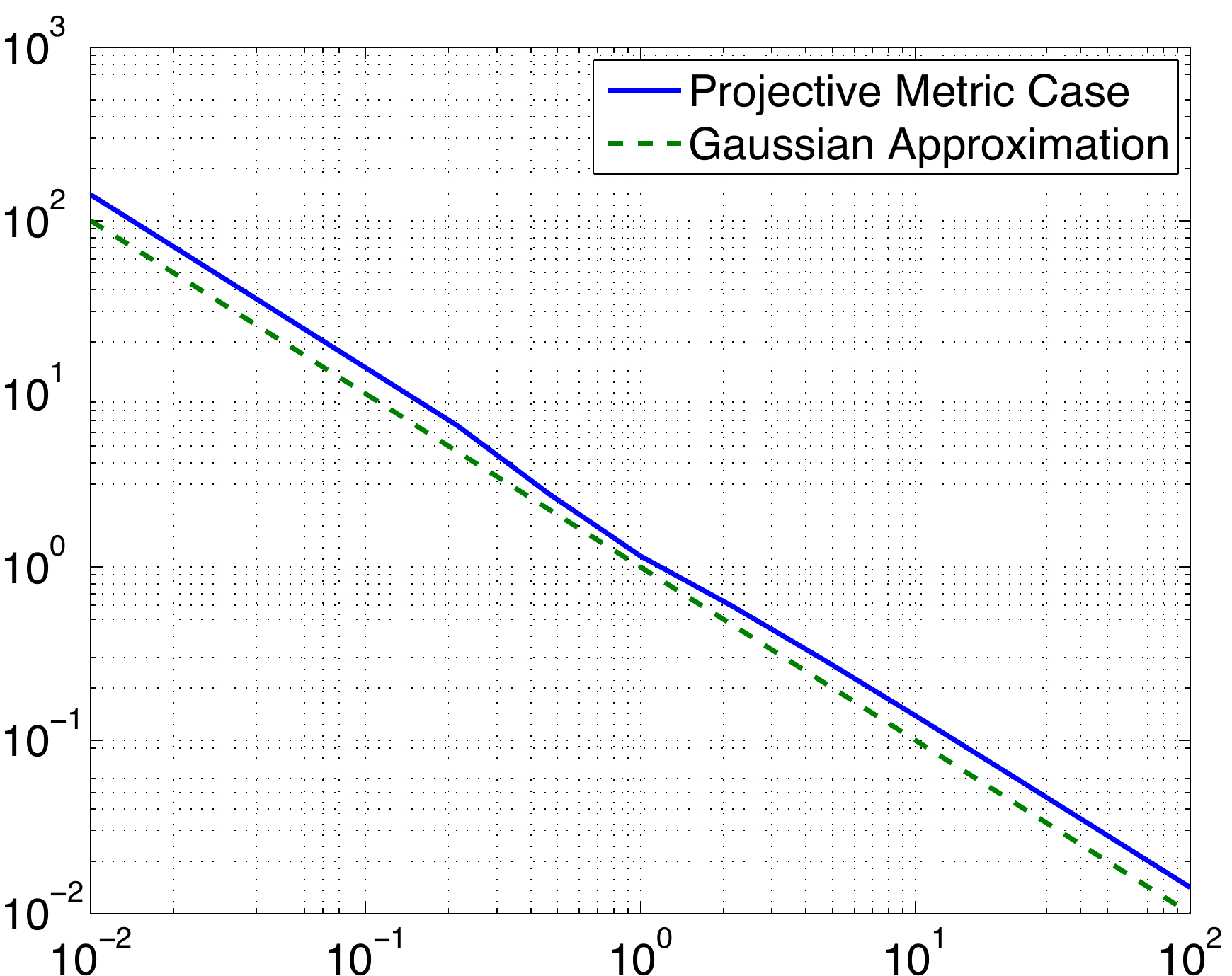}
\caption{Comparison between the prior given by (\ref{eq.prior}) and the scale invatiant
prior for the variance of a Gaussian variable.}
\label{fig:priors}
\end{figure}

In order to remove the dependence on the incoming wave angle, we consider the magnitudes of the components  $|O_n|$ instead of the signed quantities $O_n$, with the reasonable expectation  that $|O_n|$ is comparable to $|M_n|$.
Without any other intrinsic scales in the problem, the likelihood, which is non-dimensional, has to be a function of $|O_n|/|M_n|$, with a distribution that is peaked when  $|O_n|/|M_n| = 1$. Also, the observations are taken at locations that are sufficiently separated and on time scales which are sufficiently large that we can assume that they are uncorrelated. Thus, a natural choice for the likelihood is 
$$
L({\mathcal O}_n|{\mathcal M}_n,r^j_n) = \prod_{j=1}^k\frac{1}{Z^j_n} \exp\left[-\frac{1}{2(r^j_n)^2} \left(1-\frac{|O^j_n|}{|M^j_n|}\right)^2\right].
$$
where $r^j_n$ is now a dimensionless measure of the variance of $|O_n|/|M_n|$ and $Z^j_n$ is a normalization given by
\begin{equation}
Z^j_n = \sqrt{\frac{\pi}{2}} |M^j_n|r_n^j \left(1 + \mathrm{erf}\left(\frac{1}{\sqrt{2} r_n^j}\right)\right) .
\label{eq:norm}
\end{equation}
 $r^j_n$ is potentially spatially inhomogeneous (depends on $j$) and non-stationary (depends on $n$). Absent any prior information on the covariances, it is a reasonable approximation to take $r_n^j = r$, a constant. To make this more precise, we compute the Jefferys prior for $r_j^n$. The Fisher information is given by 
$$
I(r^j_n | M_n^j) = \int_{-\infty}^\infty \left[\frac{\partial \log L}{\partial r_n^j}\right]^2 L({\mathcal O}_n|{\mathcal M}_n,r^j_n) d {\mathcal O}_n,
$$
and
is independent of $M^j_n$. This is analogous to the fact that the Fisher information for the variance of a Gaussian variable with a given mean is independent of the value of the mean. We can compute the Jefferys prior $p(r_n^j) \propto \sqrt{I(r^j_n)}$, obtaining
\begin{equation}
p(r) \propto \left[\frac{e^{-\frac{1}{r^2}} \left(2 \pi  e^{\frac{1}{r^2}} r^3
   \left(\text{erf}\left(\frac{1}{\sqrt{2} r}\right)+1\right)^2+\sqrt{2 \pi }
   e^{\frac{1}{2 r^2}} \left(r^2+1\right)
   \left(\text{erfc}\left(\frac{1}{\sqrt{2} r}\right)-2\right)-2 r\right)}{\pi 
   r^5 \left(\text{erf}\left(\frac{1}{\sqrt{2} r}\right)+1\right)^2}\right]^{1/2}.
   \label{eq.prior}
\end{equation}
Figure \ref{fig:priors} shows a comparison between this prior and the scale invariant prior for the variance of a Gaussian variable, $p_G(r) \propto 1/r$. 
As is evident from the figure, to a very good approximation, we have 
$$
p(r) \approx p_G(r).
$$
We now have the Bayesian statement
$$
P(a,d,r | \mathcal{M},\mathcal{O}) \propto p(r)  \prod_{n=1}^{N} \prod_{j=1}^{k} \frac{1}{Z_n^j}\exp\left(-\frac{\Delta_{P}(1,O_n^j/M_n^j)^{2}}{2 r^{2}}\right)
{\mathcal U} ([a_{0},a_{1}]){\mathcal U}([d_{0},d_{1}]),
$$
where
\[
\Delta_{P}(f,g)=\min(|f-g|,|f+g|)
\]
is the projective metric, and $Z_n^j$ is a normalization in \eqref{eq:norm} with $r_n^j = r$.
This posterior distribution generalizes \eqref{eq.pad}. 
Note that $\Delta_P(1, O_n^j/M_n^j) = |1 - |O_n^j|/|M_n^j||$, but $\Delta_P$ has theoretically sound applications in a vectorial setup. 
In principle we can use this to estimate the various moments of the parameters $a$ and $d$ {\em and also the variance parameter} $r$ from the model output and observations.

For the purposes of this paper, we are only interested in computing the maximal likelihood estimates for $a$ and $d$ and not any of the higher moments. This allows for a further approximation which leads to further efficiencies in the computation.

The posterior distribution depends on $r$,  through (1) the prior distribution $p(r)$,  (2) through the normalization factors $Z_n$, and (3),  through the denominator in the exponent. The logarithm of the posterior distribution depends only weakly  on the first two factors. (So long as the size of $r$ be small, so that $r^{-2}$ be large compared to $\log r$). With this approximation of neglecting the $r$ dependence except in the argument of the exponential function, the posterior distribution reduces to
 \begin{equation}
P(a,d\mid\mathcal{O}) \propto \frac{1}{Z} \prod_{n=1}^{N} \prod_{j=1}^{k}\exp\left(-\frac{\Delta_{P}(1,O_n^j/M_n^j)^{2}}{2 r^{2}}\right){\mathcal U} ([a_{0},a_{1}]){\mathcal U}([d_{0},d_{1}]),
\label{eq.pad2}
\end{equation}
where $\Delta_P$ is the projective metric defined above and $Z$ is a normalization constant. In our computations, the variance parameter $r$ in the likelihood
function is set to $2$. As is evident from \eqref{eq.pad2}, changing $r$ will only change the width of the empirical likelihood, but not the value of maximum likelihood estimates.

\begin{figure}
\centering
(a)\includegraphics[scale=1,height=2.1in,width=2.4in]{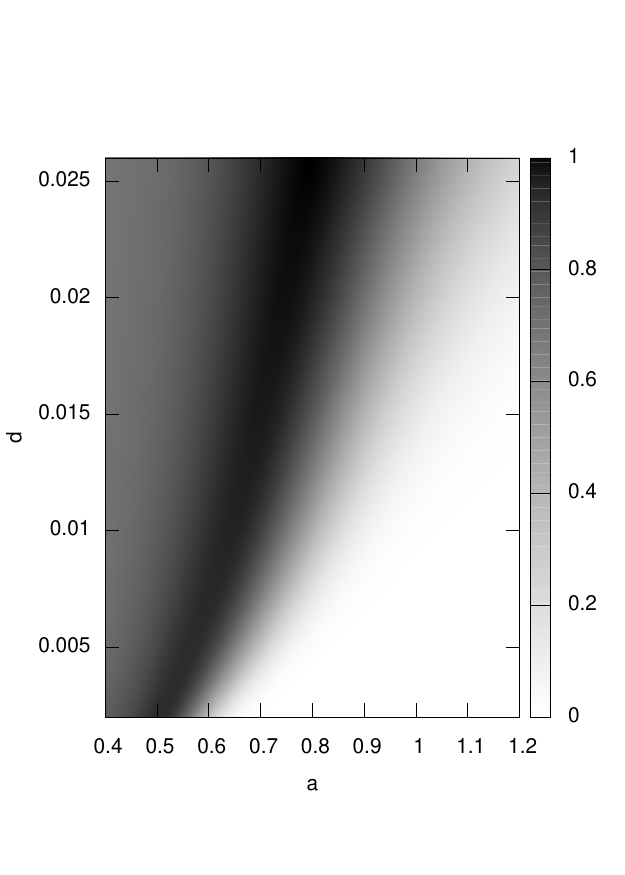}
(b)\includegraphics[scale=1,height=2.1in,width=2.4in]{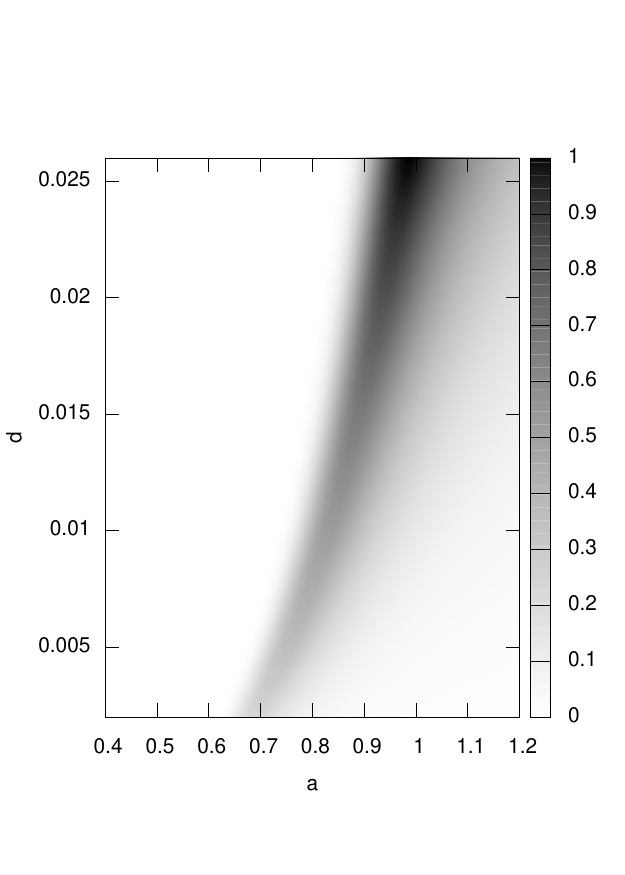}
\caption{\label{fig:oct2900} A daily comparison for October 7-8. (a)  Posterior
using absolute errors. See (\ref{eq.pad}); (b) likelihood based upon the relative errors. See (\ref{eq.pad2}).  The latter suggests that  a much sharper range of parameters  
leads to agreement between model and data.}
\label{relvsabs}
\end{figure}

Figure \ref{relvsabs}  illustrates the
difference between using absolute errors and relative errors. The striking difference
is that there is clearer discernment of likely parameter values (pure black represents the most
likely). The differences portrayed here are very drastic in this case, but this sharpening 
due to the use of the relative error is a 
generic outcome of the computations 

\subsection{Polynomial Expansion of the Parameter Sample Space}

In a  many-parameter estimation problem we would apply a Monte Carlo procedure to sample parameter space (see \cite{mchandbook}). Full model simulations make this aspect of the methodology  computationally demanding.   Whether we have to use Monte Carlo or not the efficiency of  this  process can be  considerably improved by using a parametric approximation of the model outcomes, via polynomial chaos expansions (see \cite{polychaos1}).

The polynomial chaos expansion of a stochastic function $f(\mathbf{x},t,\eta)$ of
one stochastic parameter $\eta$ is 
\[
F_{K}=\sum_{k=0}^{K}P_{k}(\eta)f_{k}(\mathbf{x},t),
\]
where $\{P_{k}\}_{k=0}^{K}$ is a system of orthogonal polynomials
of degree $k$, $0\le k\le K$. Orthogonality is with respect to the
probability distribution of the parameter $\eta$. If the distribution 
is uniform, for example,  the  $P_{k}$ are  Legendre polynomials. For two parameters,
the basis consists of polynomials $\{P_{k}(\eta)Q_{m}(\xi)\}_{k=0,m=0}^{k=K,m=M}$,
and the polynomial approximation is
\[
F_{KM}=\sum_{k=0}^{K}\sum_{m=0}^{M}P_{k}(\eta)Q_{m}(\xi)f_{km}(\mathbf{x},t).
\]
Exploiting  the orthogonality of the polynomials, we can easily find the coefficients
$f_{k_{0}m_{0}}$: We  multiply $F_{KM}$ by $P_{k_{0}}Q_{m_{0}}$
and integrate over $\eta$ and $\xi$ with respect to the known joint measure
 $\eta$ and $\xi$. In our case, this distribution
is just a constant multiple of the Lebesgue measure.

For the integration, we chose a simple Romberg integration.
However,  there are more efficient or accurate quadrature schemes, {\it e.g.}, 
Gaussian quadrature or Smolyak tensorisation (see \cite{polychaos1} and
references contained therein).

\subsection{Computational Efficiency of the Procedure}
 
We chose not to resort to Monte Carlo, since our parameter sample space is two-dimensional. Instead, we opted to discretize sample space in the range given by (\ref{eq.ranges}) with an equally spaced grid consisting of $256^2$ values.

In producing approximations to oceanic flows with numerical  models 
there is  a computational expense of ${\mathcal O}(N^d \times T)$, where $N$ is the state variable dimension, $d$ is the number of spatial dimensions (typically 2, lately 3), and $T$ is the number of time steps in the computation. $N$ is upwards of $10^5$, typically. Most codes are explicit in time, and thus $T= c \, N$ due to stability constraints, where $c$ is a constant. 

 In the numerical computations that will be described subsequently we mention that   after the initialization,
the model was run for $20$ model minutes, and the last $11$ minutes of
the model output was averaged in time and in the $y$-coordinate to
produce reference model output. 
On a 2.66 GHz dual-core machine a single model simulation took approximately  25 
wall-clock minutes to complete.

If we did not rely on the Legendre polynomial expansion, the cost of the computation
would have been $256^2 \times {\mathcal O}(N^d \times T)$. However, with the aid of 
the Legendre polynomial expansion we required 17 runs, to 
 cover the wave forcing $a$ parameter range, and 33 runs to cover the drag parameter $d$ range. 
 Thus, there were $17\times33$ full-scale runs. Once these are performed, generating 
a $256^2$ grid of parameter-space runs had a  trivial computational expense.  
In terms of wall-clock time
the Legendre calculation was about 117 times faster than the 
direct evaluation of the $256^2$ model runs.

\section{Sensitivity of the  Longshore Current with Respect to the Drag Force}
\label{analysis}

The monochromatic wave description used in the model, allows us to
relate the field observable RMS wave height, $H_{\mathrm{rms}}$,
to the wave action $W$ through the wave amplitude $A$ (by taking $H_{\mathrm{rms}} = 2A$), and $\mathbf{k}$
 the spectrum peak wavenumber. 
We use a Reynolds decomposition: For $f(x,y,t)$, we write
\[
f=\langle f\rangle+f',
\]
where $\langle f'\rangle\approx0$. The angle brackets
denote the averaging in time and in the $y$-direction.

Assuming steady waves and currents, averaging in the $x$-direction yields
\[
\langle u\rangle=-\langle u^{\mathrm{st}}\rangle,
\]
namely, the anti-Stokes condition. The mean momentum equation in the $y$-direction,
is then
\begin{equation}
\langle u\rangle\frac{\partial\langle v\rangle}{\partial x}+\langle u^{\prime}\frac{\partial v^{\prime}}{\partial x}\rangle+\langle u^{\mathrm{st}}\rangle\langle\chi\rangle
+\langle u^{\mathrm{st}\prime}\chi^{\prime}\rangle+d \langle\frac{v}{H}\rangle-\langle B_{y}\rangle=0.\label{eq:mom-y-balance}
\end{equation}
If the flow is steady, as is known to be the case for large bottom drag values, 
we obtain the balance
\begin{equation}
d \langle\frac{v}{H}\rangle=\langle B_{y}\rangle.\label{eq:Longuet-Higgins}
\end{equation}

We recall the definition of $\mathbf{B}=(B_{x},B_{y})=\frac{\epsilon_{b}\mathbf{k}}{\rho H\sigma}$,
and thus
\begin{equation}
d \langle\frac{v}{H}\rangle=\mathrm{const}\langle k_{2}\frac{H_{rms}^{7}}{H^{6}}\rangle.\label{eq:LongHig2}
\end{equation}
As remarked at the opening of the section, $H_{rms}$ can is related
to the wave amplitude in a simple and linear relation. 
The wave amplitude is determined mainly by the offshore
wave amplitude up until the breaking zone where the relatively strong
alongshore current is generated. We therefore relate $H_{rms}$ to
the offshore wave forcing $a$, a boundary-value parameter in
our numerical computations, in a linear fashion:  {\it i.e.}, we make a first
order approximation. Next we assume that the impact
of a small change in $d$ on $k_{2}$ and $H$ are negligible. This assumption is supported by the numerics. However,
when we change $d$, to satisfy the balance in (\ref{eq:LongHig2})
and assumptions leading to it, we have to change the other main parameter,
the offshore wave forcing, as well. This will change the mean longshore
velocity, as a result. We thus obtain
\begin{equation}
(d+\Delta d )\langle v+\Delta v\rangle\approx\mathrm{const}\,(H,k_{2})(H_{rms}+\Delta H_{rms})^{7},\label{eq:aim1}
\end{equation}
and 
\begin{equation}
d \langle v\rangle\approx\mathrm{const}\,(H,k_{2})H_{rms}^{7}.\label{eq:aim2}
\end{equation}
Subtracting the two relation, for small $\Delta d $ (and we assume
this results in infitesimal change to $H_{rms}$ to achieve the balance
in (\ref{eq:LongHig2})), gives
\[
\langle\Delta v\rangle\approx\mathrm{const}\,(H,k_{2},H_{rms})\frac{\Delta H_{rms}}{d+\Delta d}.
\]
We further restrict ourselves to the case where $d$ is away from
zero and relatively large compared to the increment $\Delta d$. Thus, by the aforementioned linear approximation
of $H_{rms}$ using the boundary wave forcing parameter $a$, we conclude that
\begin{equation}
\langle\Delta v\rangle\approx\mathrm{const}\,(H,k_{2},H_{rms})\frac{\Delta a}{d}.
\label{dv}
\end{equation}
The parameter estimation results that follow in the next section must reflect this dependency.

\section{Results}
\label{outcomes}

There are a variety of drag force parametrizations, but here we want to specifically test
the linear drag model. Such a model asserts that the drag force is proportional to the local depth-averaged velocity via the constant $d$. The drag force is thus only time dependent via the velocity itself. We then expect that the maximum likelihood estimation should be very stable to changes in the drag force parameters. In contrast, we can expect higher variability in the forcing amplitude. In Section \ref{analysis} we derived the structural reasons in the model that lead to this type of sensitivity.

In our study, we compare the longshore depth-averaged, time-averaged,
and spatially $y$-averaged velocity component predicted by the model
with the mean longshore current reported by Herbers,
Guza, and Elgar from experiments conducted in 1994, in Duck, North Carolina. 
The collected data, is available from 
{\tt www.frf.usace.army.mil/duck94/duck94.stm}.
The experimental
data we use here were collected by the devices v12, v13, v14, which
were located approximately at the offshore coordinates 205, 220, and
240 m, respectively.
These devices locations
cover the bar region of the domain which has stronger mean longshore
currents. 
 A plot showing the device locations as
well as a snapshot of the bottom topography appears recreated in Figure \ref{fig:topography}.

The longshore current data in this experiment was collected
at  a sampling rate of 2 Hz,  every 1024 s. There are over  5000
such data sets, spanning the months of September, October, and early
November. A small portion of the data was not used, either
because there was a device failure or because the  datum was an 
extreme outlier.
Since there were several experimental devices with approximately
the same longshore coordinate, model output as well as instrumental
data was $y$-averaged. As a result, the model-data discrepancy has
no longshore dependence. 
The month of September, October, and November observations include 5030 data points for the longshore mean current.
First 2370 data points come from the observations in September 1994, the next 2450 from October 1994, 
the last 210 in November 1994. In the text, we number these data points consecutively.

The general outcomes can be summarized as follows: 
\begin{itemize}
\item The proposed likelihood was superior to the more traditional, absolute distance
likelihood, given that there was no covariance information available. In particular, it delivered sharper maximally-likely parameter values for model/data agreement.
\item With regard to the polynomial chaos expansion, we found that  the first few terms in the sequence were 
of significance  in all the cases considered: 
 For  the wave forcing amplitude $a$
we use the first $6$ coefficients, whereas for the drag coefficient
$d$  we used the first $9$ coefficients. We found that using more coefficients 
did not significantly improve the results.
\item The use of the polynomial chaos expansion improved the efficiency of the parameter
estimation by 2 orders of magnitude.
\item We
found that the  most likely  wave forcing was in the range, $a=0.8-1.1$ m.  The
most likely bottom drag coefficient was in the range $d=0.007-0.020$. The model
is thus most consistent with the data when wave amplitudes are large, but
not excessively so, and in the range of dynamics wherein the attractor of
solutions reflects a balance of breaking and drag forces and near-steady
longshore currents. 
\item With regard to the length of the experiments, we found that a minimum of
3-5 hours of data were needed. Among the  reasons for this is  that below 3 hours
the number of time records were too few: less than twelve. 
\item When several 3 hour data
experiments were compared, we found  variations in
the wave forcing estimate. 
This is a positive modeling outcome: variations in the wave forcing are tied to the time scales of wind variation, which is roughly 3-6 hours. Figure \ref{fig:threehour-1} is typical
of estimates at different times of the day. The times of these observations are September 25, 16:51-19:51 and September 25-26, 23:59-3:16.
\begin{figure}
\centering
(a)\includegraphics[width=2.1in,height=2in]{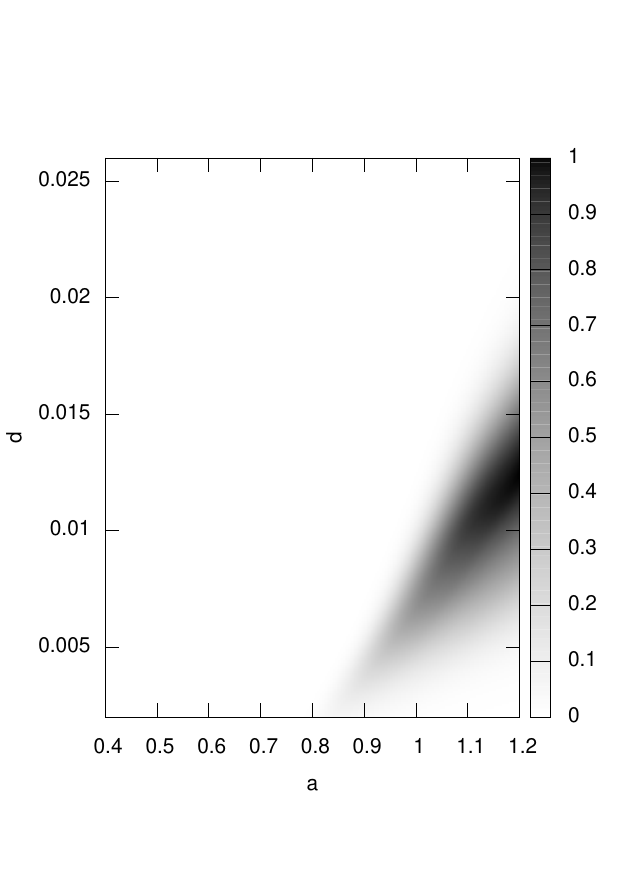}
(b)\includegraphics[width=2.1in,height=2in]{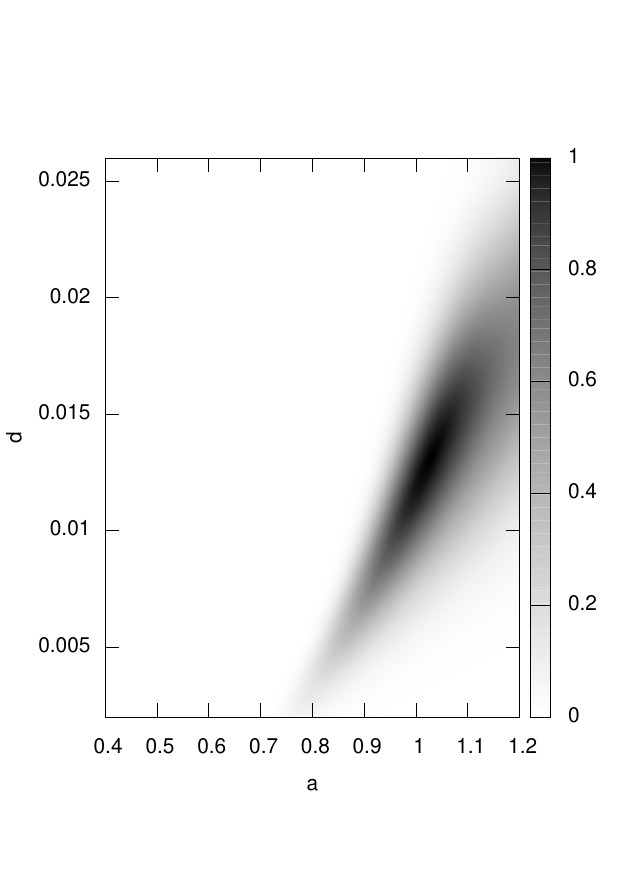}
\caption{Three-hour comparisons of data and model output. 
(a) Posterior for data points 1944 through 1955;  a comparison performed starting
3 hours later, (b) another three-hour posterior for data points 1968 through 1979. 
We note that the wave forcing changes significantly, but not the drag. Nevertheless,
the results are reasonable, and the methodology permits tracking this change
in parameter estimates.} 
\label{fig:threehour-1}
\end{figure}
\item  On the other hand, we found that the drag coefficient was not as sensitive to 
the length of the experiment: Whether using 3 hour or daily data, the results were similar, with relatively high bottom drag coefficients favored; nevertheless, 
slightly larger than 0.007, the marginally stable value used in \cite{UMR}.
For the daily runs,   partitions of 96 consecutive data points in the observations were used. In Figure \ref{fig:med-daily}  we show a ''daily'' case, from September 1. 
\begin{figure}
\centering
(a)\includegraphics[width=2.in,height=1.9in]{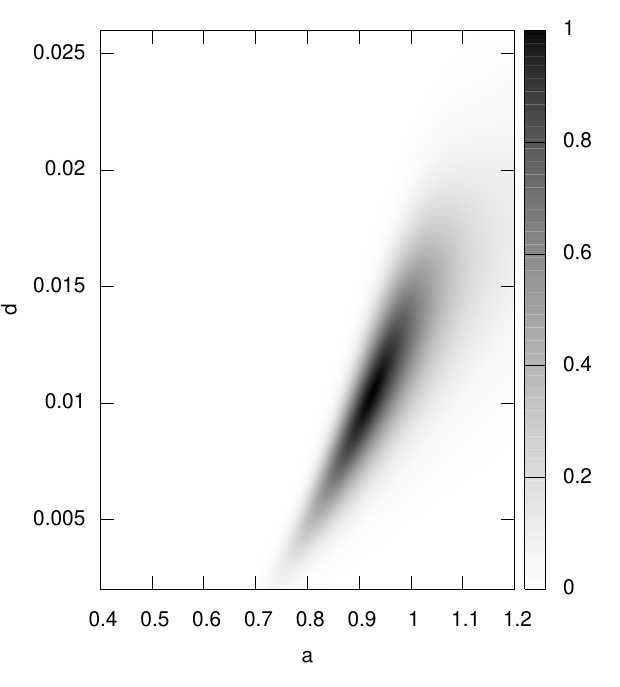}
(b)\includegraphics[width=2.2in,height=1.85in]{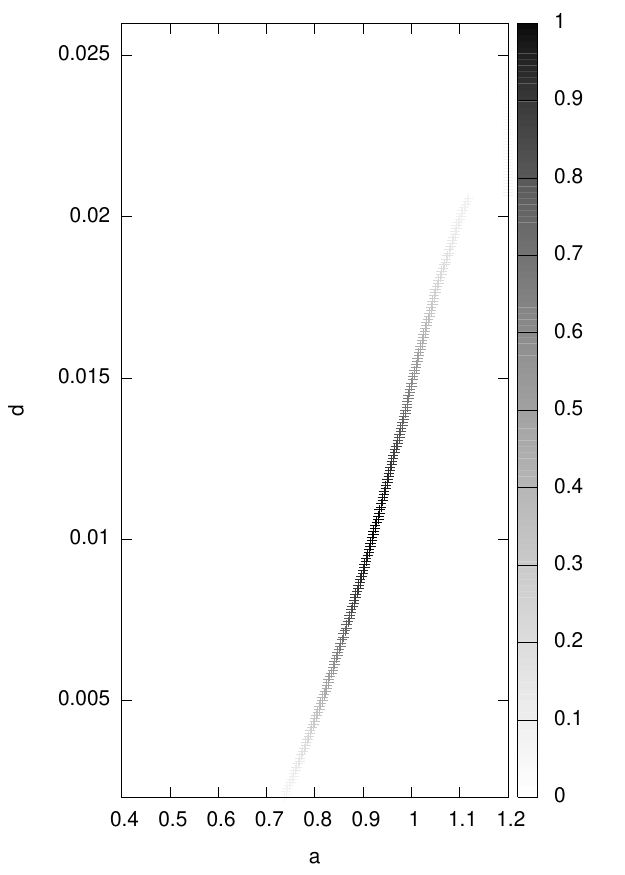}
\caption{Relatively stronger wave heights induce similar
daily posteriors with more spread out modes. We note the steep curve traced by
the most likely parameter values, with large variability in the bottom drag, and 
less variability in the forcing amplitude. The maximum likelihood curve in parameter space, associated with 
the likelihood function is shown in Figure \ref{fig:med-daily}b. The estimate in (\ref{eq:nusretslaw}) qualitatively conforms with the curve.}
\label{fig:med-daily}
\end{figure}
\item 
Tracing the maximum likelihood curve in $a\times d$ parameter space (see Figure \ref{fig:med-daily}) we obtain
a curve, which when combined with 
(\ref{dv}), recovers a semi-empirical law which holds for our case of interest:
\begin{equation}
\langle\Delta v\rangle\sim_{H,k_{2},H_{\mathrm{rms}}}\frac{\Delta d}{d}.\label{eq:nusretslaw}
\end{equation}

In obtaining this relationship we assumed that
$\Delta H_{rms}\approx\mathrm{const}\,\times\Delta d$. This semi-empirical 
relationship says that 
 the mean longshore velocity in the balance (\ref{eq:LongHig2})
depends on the drag parameter in a locally logarithmic manner. This
is consistent with the shapes of the posterior distributions,
and the numerical observations in the sample runs: the higher the
drag, the lesser the impact of the change in the drag on the velocity
field. This is particularly evident in Figure \ref{fig:med-daily}b, which
highlights the maximum likelihood curve in parameter space, corresponding to
data points 700 through 795.
\end{itemize}

\begin{figure}
\centering
\includegraphics[scale=1,height=2.1in,width=2.8in]{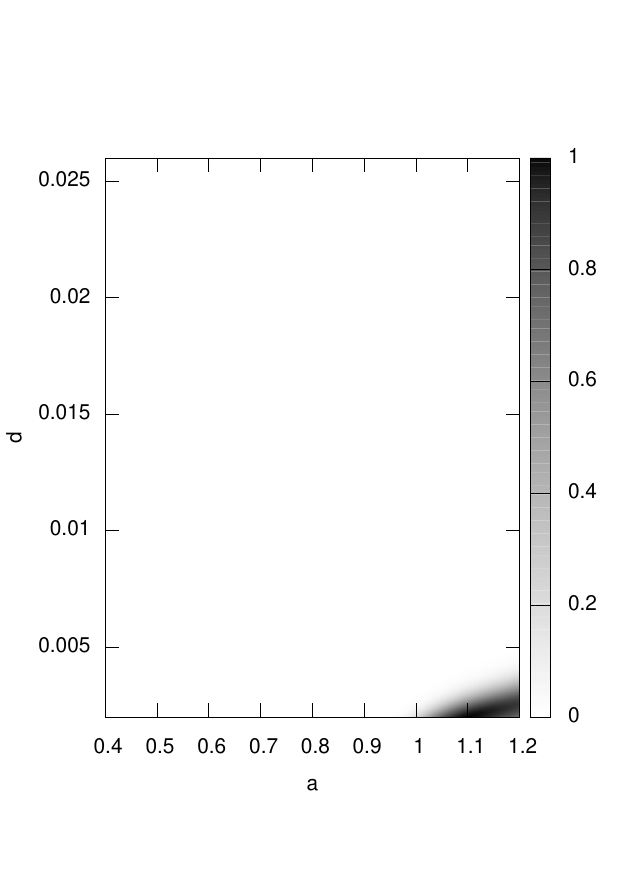}
\caption{The Bayesian 
posterior is at odds with what we know about the physics of longshore currents.
Data points 3800 through 3895 are characterized by the presence of very strong currents. The model is unable to capture the data.}
\label{fig:too-strong-daily}
\end{figure}

The tuning strategy is not failure-proof, but this is not seen here in a negative
light: If the model and the data are irreconcilable it is possible for one
to obtain estimates that make little sense. 
That we know they do not
make sense means that we know something about the physics and
limitations of the model for us to make this determination. We do not have this much knowledge about the outcomes in other complex problems.
For models that are exceedingly complicated this becomes a nontrivial
challenge. Nevertheless, 
priors can be used to narrow the parameter search or for constraining
its characteristics: More priors can be embedded into (\ref{eq.pad2}). 
The danger, however, is that these priors inform the posterior too
strongly; what is desired is that the priors inform the likelihood.
A non-sensical case is shown in Figure \ref{fig:too-strong-daily}, using
a data set corresponding to measurements 3800 through 3895.
This is a daily-data experiment
 suggesting that the likely drag values are exceedingly small and the wave forcing 
 very high.
Model outcomes corresponding to this case would correspond to highly unstable
modeled longshore currents with variability that is not consistent with measurements.

\section{Summary}

We proposed an alternative maximum likelihood function for model/data disparity for
parameter estimation, suitable in cases when covariance information is unavailable or 
poorly constrained. The lack of this information  frequently  occurs in real-world applications 
and thus it is  argued that this alternative likelihood function will have practical value.

The covariance matrix not only provides weights for model/data disparity. 
 It is also crucial in nondimensionalizing and setting properly the scales of the 
 state vector.
If the state vector is composed of multi-physics components, the lack of scale information might become an unsurmountable challenge in practical parameter estimation. One could non-dimensionalize the state vector in the model and the data, but doing so in a large scale model (or in an existing code that approximates solutions to the model) can be impractical.
 Moreover, even if there is no multi-physics, but rather, a multiplicity of 
spatio-temporal scales, the challenge will not be easily obviated by a fixed scaling.

With regard to parameter estimation we showed that using the traditional maximum likelihood variant (presuming no covariance information is known), produces parameter estimates that can be vague, or at worst, uninformative. 
The alternative maximum likelihood function replaces the weighted model/data disparity, the absolute error, by a relative error between these. In doing so we make sure that 
the likelihood function will automatically adjust in a reasonable way its variance according to the scales of the state vector components (and of its spatio-temporal characteristics).   The use of the relative error will make larger demands with regard to the compatibility of the observations and the model based upon the importance, in terms of magnitude, of the model state vector. This leads to sharper estimates. This is not generally the case when the absolute error is used:  discrepancies between model output and observations are not ordered with regard to their importance, unless judiciously done so in an ad-hoc fashion. 
A likelihood which is cast in terms of the relative error changes the issue of the relative confidence of the individual elements of the absolute error to one in terms of relative error changes. That is, one is now required to know the dimensionless measure of the variance of 
the relative size of measurements to model outcomes. Under reasonable assumptions the Jeffreys prior can be used to justify the replacement of each dimensionless measure of the variance by a single constant.

Large-scale dynamics models make parameter estimation extremely challenging, due to the curse of dimensionality: firstly, because of the combinatoric complexity of choosing parameters within a model, and secondly, because the models tend to have many degrees of freedom.    In our case, we only had 2 parameters and thus we did not have to resort to Monte Carlo and experimental design. For the latter, 
the adoption of a Legendre polynomial approximation to the model output  increased the efficiency of parameter estimation  one-hundred-fold.  

We also derived an estimate of the parameter dependence, between the drag and the wave forcing; namely, the rate of variability of the model outcomes to changes in these two parameters (see (\ref{dv})). We used this estimate to verify  that the 
likelihood estimates, derived using our new formulation were qualitatively correct. 

  The longshore problem has at least two qualitatively different flow regimes. According to the model, the model tracks the Duck data in a flow regime wherein 
  the drag forces and the breaking wave forces are in a near-balance and the longshore current nearly steady; in that regime we found  likely parameter combinations. The sensitivity of the parameters in the estimation reflected well a theoretical estimate of their relative
  variability, derived from the model itself. The  linear drag model was validated in this steady-longshore flow regime. It should be noted that the data did not lend itself to 
  evaluating the drag model in the   more    time-unstable flow regime.

\section*{Acknowledgments}

We received funding from GoMRI/BP. JR and SV also  received funding from 
NSF-DMS-1109856. 
This work was supported in part by the National Science Foundation under Grant No. PHYS-1066293 and the hospitality of the Aspen Center for Physics.  JR wishes to thank Andrew Stuart, for stimulating discussions, as well as the American Institute of Mathematics.

%
%

\begin{thebibliography}{17}
\providecommand{\natexlab}[1]{#1}
\providecommand{\url}[1]{\texttt{#1}}
\expandafter\ifx\csname urlstyle\endcsname\relax
  \providecommand{\doi}[1]{doi: #1}\else
  \providecommand{\doi}{doi: \begingroup \urlstyle{rm}\Url}\fi

\bibitem[Alexanderian et~al.(2012)Alexanderian, Winokur, Sraj, Srinivasan,
  Iskandarani, Thacker, and Knio]{polychaos1}
Alen Alexanderian, Justin Winokur, Ihab Sraj, Ashwanth Srinivasan, Mohamed
  Iskandarani, William~C. Thacker, and Omar~M. Knio.
\newblock Global sensitivity analysis in an ocean general circulation model: a
  sparse spectral projection approach.
\newblock \emph{Computational Geoscience}, 16:\penalty0 757--778, 2012.
\newblock \doi{10.1007/s10596-012-9286-2}.

\bibitem[Elgar et~al.(1994)Elgar, Herbers, and Guza]{elgar1994}
S.~Elgar, T.~H.~C. Herbers, and R.~T. Guza.
\newblock Reflection of ocean surface gravity waves from a natural beach.
\newblock \emph{Journal of Physical Oceanography}, 24:\penalty0 1503--1511,
  1994.

\bibitem[Eyink et~al.(2004)Eyink, Restrepo, and Alexander]{ERA-I}
G.~L. Eyink, J.~M. Restrepo, and F.~J. Alexander.
\newblock A mean field approximation in data assimilation for nonlinear
  dynamics.
\newblock \emph{Physica D}, 194:\penalty0 347--368, 2004.

\bibitem[Kroese et~al.(2011)Kroese, Taimre, and Botev]{mchandbook}
D.~P. Kroese, T.~Taimre, and Z.~I. Botev.
\newblock \emph{Handbook of Monte Carlo Methods}.
\newblock John Wiley and Sons, New York, 2011.

\bibitem[Lane et~al.(2007)Lane, Restrepo, and McWilliams]{lrm06}
E.~M. Lane, J.~M. Restrepo, and J.~C. McWilliams.
\newblock Wave-current interaction: A comparison of radiation-stress and
  vortex-force representations.
\newblock \emph{Journal of Physical Oceanography}, 37:\penalty0 1122--1141,
  2007.

\bibitem[McWilliams and Restrepo(1999)]{MR99}
J.~C. McWilliams and J.~M. Restrepo.
\newblock The wave-driven ocean circulation.
\newblock \emph{Journal of Physical Oceanography}, 29:\penalty0 2523--2540,
  1999.

\bibitem[McWilliams et~al.(2004)McWilliams, Restrepo, and Lane]{mrl04}
J.~C. McWilliams, J.~M. Restrepo, and E.~M. Lane.
\newblock An asymptotic theory for the interaction of waves and currents in
  coastal waters.
\newblock \emph{Journal of Fluid Mechanics}, 511:\penalty0 135--178, 2004.

\bibitem[Restrepo(2008)]{drifter}
J.~M. Restrepo.
\newblock A path integral method for data assimilation.
\newblock \emph{Physica D}, 237:\penalty0 14--27, 2008.

\bibitem[Sepahvand et~al.(2010)Sepahvand, Marburg, and Hardtke]{basicpc}
K.~Sepahvand, S.~Marburg, and H.~J. Hardtke.
\newblock Uncertainty quantification in stochastic systems using polynomial
  chaos expansion.
\newblock \emph{International Journal of Applied Mechanics}, 2:\penalty0
  305--353, 2010.

\bibitem[Severini(2001)]{Loglike}
Thomas~A. Severini.
\newblock \emph{Likelihood Methods in Statistics}.
\newblock Oxford University Press, New York, 2001.

\bibitem[Sraj et~al.(2013)Sraj, Winokur, Alexanderian, Lee, Chen, Srinivasan,
  Iskandarani, Thacker, and Knio]{polychaos2}
Ihab Sraj, Justin Winokur, Alen Alexanderian, Chia-Ying Lee, Shuyi~S. Chen,
  Ashwanth Srinivasan, Mohamed Iskandarani, William~C. Thacker, and Omar~M.
  Knio.
\newblock {B}ayesian inference of wind drag dependence on wind speed 2 using
  {AXBT} data during {T}yphoon {F}anapi.
\newblock \emph{Ocean Modeling}, 2013.

\bibitem[Stuart et~al.(2004)Stuart, Voss, and Wiberg]{voss}
A.M. Stuart, J.~Voss, and P.~Wiberg.
\newblock Conditional path sampling of {SDE}s and the {L}angevin {MCMC} method.
\newblock \emph{Communications in the Mathematical Sciences}, 2:\penalty0
  685--697, 2004.

\bibitem[Thornton and Guza(1983)]{tgbreaking}
E.~B. Thornton and R.~T. Guza.
\newblock Transformation of wave height distribution.
\newblock \emph{Journal of Geophysical Research}, 88 (C10):\penalty0
  5925--5938, 1983.

\bibitem[Uchiyama et~al.(2009)Uchiyama, McWilliams, and Restrepo]{UMR}
Y.~Uchiyama, J.~C. McWilliams, and J.~M. Restrepo.
\newblock Wave-current interaction in nearshore shear instability analyzed with
  a {V}ortex {F}orce formalism.
\newblock \emph{Journal of Geophysical Research}, page C06021, 2009.
\newblock \doi{doi:10.1029/2008JC005135}.

\bibitem[Weir et~al.(2011)Weir, Uchiyama, Lane, Restrepo, and
  McWilliams]{WULRM}
B.~Weir, Y.~Uchiyama, E.~Lane, J.~M. Restrepo, and J.~C. McWilliams.
\newblock A {V}ortex {F}orce analysis of the interaction of rip currents and
  surface gravity waves.
\newblock \emph{Journal of Geophysical Research}, 116:\penalty0 C050001, 2011.

\bibitem[Wiener(1938)]{pcwiener}
N.~Wiener.
\newblock The homogeneous chaos.
\newblock \emph{American Journal of Mathematics}, pages 897--936, 1938.

\bibitem[Wunsch(1996)]{wunschbook}
C.~Wunsch.
\newblock \emph{The Ocean Circulation Inverse Problem}.
\newblock Cambridge University Press, Cambridge, UK, 1996.

\end{thebibliography}
%

\end{document}